# Real-Time, Label-free Electrical Transduction of Catalytic Events in a Single-Protein Redox Enzymatic Junction

Tracy Quynh Ha[1], Albert C. Aragonès[2*], Qiankun Wang[1], Desmond Koomson[1], Nashili Kibria[1], Jhanelle White[1], Kavita Garg[1,3], Jessica Peate[1], Alex P.S. Brogan[1], Leigh Aldous[1,4], Sarah M. Barry[1*], Ismael Díez-Pérez[1*]

[1]Department of Chemistry, Faculty of Natural, Mathematical & Engineering Sciences, King's College London, Britannia House, 7 Trinity Street, London SE1 1DB, UK.
[2]Department of Physical Chemistry, Max Planck Institute Partner Group, University of Barcelona, Martí i Franquès 1, 08028, Barcelona, Spain.
[3]Department of Chemistry, School of Physical Sciences, DIT University, Dehradun-248001.
[4]Department of Chemical Engineering, National Taiwan University, Roosevelt Road, Taipei 106, Taiwan.

***e-mail**: ismael.diez_perez@kcl.ac.uk, sarah.barry@kcl.ac.uk, acortijos@ub.edu

## Abstract

Understanding redox enzymatic reactions remains a fundamental challenge in biochemistry, as these processes involve intricate redox and structural dynamics that are inherently difficult to dissect using conventional bulk techniques. Single-enzyme catalysis offers a promising approach for unravelling the dynamic behaviour of individual enzymes as they undergo a reaction, revealing the complex heterogeneity that is lost in the averaged ensembles. Here we demonstrate real-time, label-free monitoring of the electrical transduction of single-protein enzymatic activity for two redox enzymes, cytochrome $P450_{cam}$ and glutathione reductase, trapped in an electrochemically controlled nanoscale tunnelling junction immersed in the aqueous enzymatic mixture. The conductance switching signal observed in individual transients of the electrical current flowing through the single-protein junction shows that the tunnelling conductance is modulated by the enzymatic reaction; subtle changes of the enzyme redox state occurring during the chemical catalysis process result in fluctuations of the enzyme junction conductivity, which are captured as a switching signal. At the applied electrochemical reducing potential for electrocatalysis, the transient oxidation of the trapped enzyme in every catalytic cycle opens an additional redox-mediated electron tunnelling channel in the single protein junction that results in a temporary current jump, contributing to the observed conductance switching features. The latter is experimentally assessed via electrochemically controlled conductance measurements of the single-protein junction. The statistical analysis of the switching events captured over long time periods results in average frequencies that correlate well with the reported catalytic turnover values of both enzymes obtained in standard bulk assays. The single-enzyme experiments reveal the acute heterogenous behaviour of enzymatic catalysis and the quantification of single enzyme turnover frequencies. The work demonstrates a new nanoscale platform for non-labelled electrical detection of single-enzyme activity.

# Main

Redox enzymes are known for their ability to catalyse kinetically demanding chemical reactions in many cellular processes[1]. Despite the vast crystallographic information of enzyme structure, we have still very little understanding about the enzyme's outstanding ability to accelerate chemical reactions, in some instances, million times faster than we can achieve by synthetic means, and with remarkable stereo- and regio-specificity. Such remarkable catalytic power has led to extensive efforts to utilise enzymes as biocatalysts in the chemical synthesis of chiral building blocks and high value-added products such as drugs[2,3]. Understanding the fundamental mechanism of enzymatic catalysis is also key to inform protein engineering towards improved enzymes for practical biotechnology. Traditional ensemble methods investigating enzymatic activity generally provide averaged results, which does not reflect intracellular environments that exhibit high heterogeneity[4]. Single-protein junctions have brought new prospects into single-enzyme studies[5,6,7,8], enabling high-resolution and high-throughput single-protein electrical characterization to access complex enzymatic reaction mechanisms[6]. Scanning tunnelling microscopy (STM)-based molecular tunnelling junctions have proven to be very sensitive to the oxidation state of the molecule. Early experiments by Tao demonstrated single-molecule visualisation of redox-dependent current enhancement on a redox *heme* co-factor using an electrochemical STM (EC-STM)[9]. Ulstrup and Kuznetsov later demonstrated the same effect on a redox Cu-azurin protein, showing that the tunnelling currents flowing through an individual redox protein are strongly regulated by the redox state of the protein metal cofactor[10,11]. Following this pioneering work, we demonstrated *in situ* real-time dynamic detection of transient redox changes in an engineered Cu-Azurin trapped in an EC-STM-based nanogap[7]. The latter opens to a whole new vista in single-protein detection of redox enzymatic catalysis, as the redox enzymatic process of a single trapped enzyme in a tunnelling junction can be followed via dynamical changes in redox state occurring during the catalytic cycle. This vision has been long predicted by Albrecht, who proposed that the tunnelling conductance of an active trapped enzyme in a nanogap would produce a discrete conductance switching signal generated by individual catalytic turnover events[12]. Similar switching behaviour has been observed in previous single protein fluorescence microscopy studies. For example, Lu *et al.* investigated the catalytic process of a FAD-dependent redox enzyme, cholesterol oxidase (COx)[13]. They observe that the COx fluorescence intensity fluctuates in the presence of its enzymatic substrate, and describe the phenomenon as on-off behaviour resulting from enzymatic turnover when FAD toggles between being oxidised (on state) and reduced (off state)[13]. While fluorescence detection has been proven to reach single-enzyme detection levels[14,15,16,17], its broad applicability is challenging as (1) it requires fluorescent labels needing enzyme modification, which limits the scope of enzymes that can be studied, and (2) they are prone to photobleaching and pH dependences[18], limiting long-term studies and throughput. Very recent pioneering work in the area of single-protein electronics has also shown that tunnelling current noise produced by a single non-redox Φ29 polymerase trapped in a nanogap can be ascribed to enzymatic activity[5]. Zhang *et al.* also exploited single-protein junctions to capture intermedium states of the catalytic cycle in the enzyme formate dehydrogenase, proving that different catalytic states can be ascribed to different protein conductance states[6].

Herein, we combine all the above pioneering efforts in redox single-protein junctions to develop a new methodology for real-time, label-free, single-protein electrical transduction of redox enzymatic catalysis. We demonstrate the methodology by studying two very different unmodified redox enzymes, namely, cytochrome P450$_{cam}$ and glutathione reductase (GR), with contrasting active redox cofactors *heme* and flavin adenine dinucleuotide (FAD), respectively, catalysing very different chemistries at very disparate catalytic rates. Under catalytic conditions, individual trapped enzymes in an electrochemically controlled nanogap at a fixed potential generate distinct two-level tunnelling conductance fluctuations matching the ones experimentally observed for the same single-enzyme junctions under varying electrochemical potentials and non-active catalytic conditions. The characteristic frequency of the observed conductance switching is directly compared against the reported enzymes' catalytic rates by collecting thousands of single-protein catalytic events and averaging them over the total residence time of the trapped proteins in the nanogap. The analysis shows an exquisite agreement between the observed single-enzyme electrical fluctuations collected in long time periods and the bulk enzyme catalytic rates of the two enzymes. The latter validation opens to a new methodology for real-time, label-free, electrical detection of catalytic events in an individual redox enzyme, enabling (1) the exact quantification of catalytic heterogeneity (ON-OFF periods) in redox enzymatic processes, (2) the real single-enzyme catalytic frequencies, and (3) electronic sensing platforms for high-resolution analytical quantification of enzymatic activity.

## Results

### Preparation of Enzymes/Electrode Interfaces for Single Enzyme Entrapment

Cytochrome P450 enzymes are a ubiquitous superfamily of *heme* dependent enzymes with vital roles in many biochemical pathways, such as hormone biosynthesis, and have long been important in drug development due to their role in xenobiotic metabolism[2,3,19]. As we attempt to transition to more sustainable chemical processes, P450s have gained importance as biocatalysts[19,20]. This is due to their ability to catalyse a wide variety of synthetically challenging reactions such as hydroxylation, epoxidation and C-C bond formation via regio and enantioselective activation of inert C-H bonds, using molecular oxygen as a co-substrate[21]. In this work, we focus on the well-characterised cytochrome P450$_{cam}$, which catalyses the kinetically difficult hydroxylation of D-camphor to 5-*exo*-hydroxycamphor at modest catalytic rates ($k_{cat}$) of 124-3960 min$^{-1}$ [22,23,24], and compare it to a GR enzyme, which catalyses the less complex reduction of oxidised glutathione (GSSG) to reduced glutathione (GSH) at a much higher $k_{cat}$ of 12600-17500 min$^{-1}$ [25,26,27] (see comprehensive catalytic cycles in Supplementary Fig. S24-25).

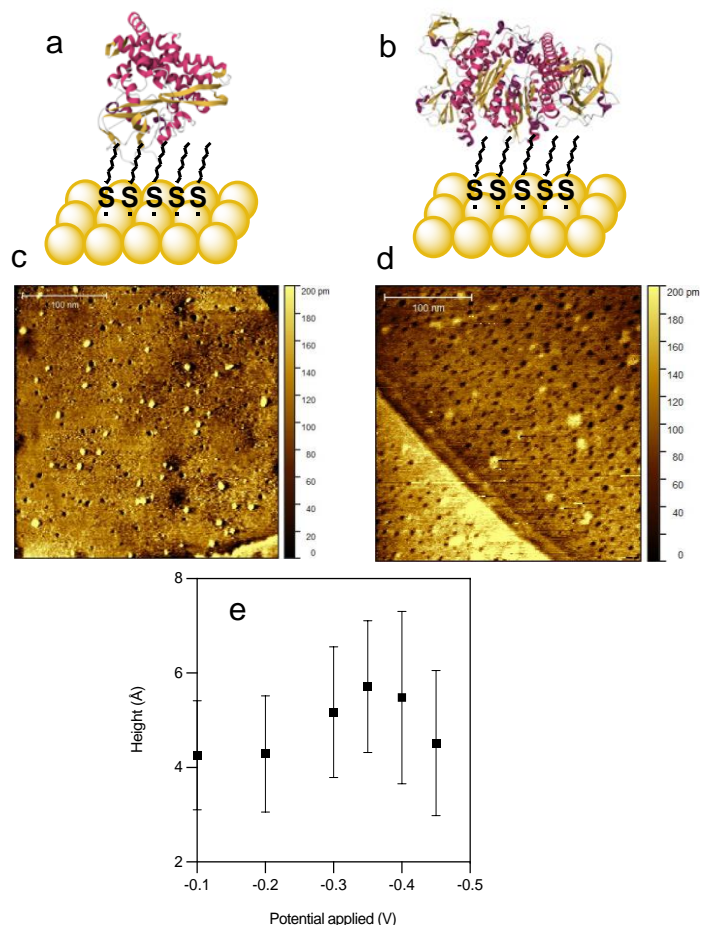

**Fig. 1**. **EC-STM imaging of P450$_{cam}$-C8-Au and GR-C8-Au**. **a,b,** Schematic representation of both P450cam-C8-Au and GR-C8-Au interfaces, respectively. **c,d,** EC-STM images of P450$_{cam}$ in 0.1 M phosphate buffer (pH 7) and GR in 0.1 M Tris-KCl (pH 7), respectively, immobilised on 1-octanethiol functionalised Au(111), with the Au(111) subjected to an electrochemical potential of 0.1 V vs AgCl reference electrode. **e,** Average STM apparent height of individual P450$_{cam}$ as a function of the applied electrochemical potential. All images were collected with a Au(111) working electrode, Pt counter electrode and a Ag/AgCl (3 M KCl) reference electrode. $V_{bias}$ = +0.1 V, current setpoint = 1 nA.

Unmodified recombinant P450$_{cam}$ or GR enzymes were adsorbed via non-specific van der Waal interactions on an atomically flat Au(111) electrode surface functionalised with a 1-octanethiol self-assembled monolayer (SAM) providing a soft electrode/protein contact. The interfaces are denoted as P450$_{cam}$-C8-Au (Fig. 1a) and GR-C8-Au (Fig. 1b), respectively for the two enzymes. The two interfaces were imaged at high resolution using an EC-STM (Fig. 1c and d, respectively) under potentiostatic conditions in 0.1 M phosphate (pH 7) and 0.1 M Tris-KCl buffers (pH 7) for P450$_{cam}$ and GR, respectively. The dark pits observed in the image background are indicative of the formation of a uniform octanethiol SAM[28]. Sitting on top of the SAM, the enzymes appear as brighter spots, which are stable upon successive STM image scans showing successful enzyme adsorption[11,29,30]. The average apparent diameters of both enzymes in the STM images (Supplementary Fig. S16) result in average diameters of 5.4 and 10 nm for P450$_{cam}$ and GR, respectively, consistent with the differences in the crystallographic size between the two proteins (Supplementary Fig. S14). The slight enzyme-to-enzyme contrast variability of the proteins in the STM images denotes small variations in their orientations on the electrode surface together with the conductivity mapping nature of the STM images which tends to enhance the more conductive areas around the redox cofactor[9]. The latter is also responsible for the observed anomalously small protein apparent heights in the STM image, typically of only several Å, which reflect their lower conductance as compared to the electrode substrate underneath[9,11,30]. The limited spatial resolution of the topographical images in tapping mode atomic force microscopy (AFM) allowed imaging the large GR only (Supplementary Fig. S15), measuring average protein heights of 6.5 nm and suggesting the enzymes are generally "lying down" on the electrode surface (Supplementary Fig. S14). To further corroborate the STM images of individual proteins, we have conducted electrochemical potential-dependent STM imaging of the P450$_{cam}$-C8-Au. The STM imaging contrast in redox molecules depends on the energy alignment of the redox state with the Fermi energy of both STM electrodes, which enhances tunnelling probability when in resonance and results in higher apparent heights. This has been demonstrated in several redox proteins[7,11,31], including enzymes[32]. We observe an abrupt increase in the apparent STM height of P450$_{cam}$ at the enzyme redox midpoint potential -0.35 V (Fig. 1e), in accordance with the cyclic voltammetry (CV) results (Fig. 3a), which reinforces the assignment of the bright spots in STM images to individual redox enzymes.

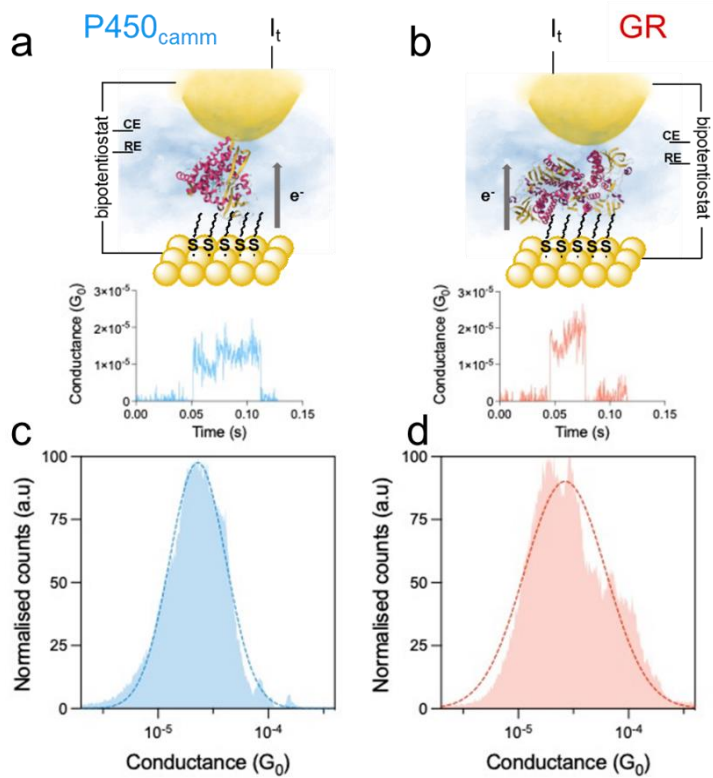

**Fig. 2**. **Single-protein conductance measurements. a**,**b,** Top panels illustrate the experimental setup of the trapped enzyme junctions, and bottom panel represents their corresponding current-time trace for P450$_{cam}$-C8-Au and GR-C8-Au, respectively. **c**,**d,** 1D conductance histograms of accumulated current-time measurements for single for P450$_{cam}$ and GR junctions, obtaining average conductance values of 2.2 x 10$^{-5}$ G$_0$ and 2.6 x 10$^{-5}$ G$_0$, respectively. EC-STM current-time measurements were obtained at an electrochemical potential = +0.1 V, V$_{bias}$ = +0.1 V, low current setpoints = 40 pA with Au(111) and STM probe as the working electrodes, Pt as the counter electrode and Ag/AgCl (3 M KCl) as the reference electrode. Each histogram accumulates ~3200 – 4000 traces.

Next, we trap individual enzymes in a static, electrochemically controlled nanoscale gap as follows: the STM tip electrode is brought into proximity to the thiol-modified Au(111) substrate, creating gap distances ranging 4.0 – 5.7 nm for GR and 3.8 – 5.2 nm for P450$_{cam}$ (gap distance calibration details in Supplementary Fig. S13). The largest distances were achieved by setting a small tunnelling current set point (a few tens of pico-Ampers) at an applied bias voltage difference (V$_{bias}$= U$_{sample}$ − U$_{tip}$) of +0.1 V between the two junction electrodes. Once the STM tip-to-substrate gap is mechanically stable, the current feedback loop is deactivated, and the tunnelling current continuously monitored during the lifetime of stability of the tunnelling nanogap (typically <1 sec). During this time, spontaneous enzyme trapping events between the two Au electrodes, STM probe and Au substrate (Fig. 2a-b top panels), lead to abrupt increase in the measured tunnelling current, which manifests as distinct '*jumps''* or "*blinks*" of short duration (~10-100 msec), as shown in the lower panels of Fig. 2a-b for P450$_{cam}$ and GR, respectively. Thousands of these blinks were accumulated without data selection in 1D conductance histograms (Fig. 2c-d, respectively for P450 and GR), and using Gaussian distribution functions, we yield average conductance values for the two proteins: $2.2 \times 10^{-5}$ $G_0$ and $2.6 \times 10^{-5}$ $G_0$ for P450$_{cam}$ and GR, respectively ($G_0 = 2e^2/h \approx 77.4$ μS). These conductance values are in good agreement with the conductance previously measured for redox proteins of comparable size[7,33].

## Electrocatalytic Activity of the P450cam-C8-Au and GR-C8-Au Interfaces

The electrocatalysis of both the P450-C8-Au and GR-C8-Au interfaces was probed electrochemically and the electroactivity confirmed by analysing the extracted electrocatalytic reaction buffer using gas chromatography mass spectrometry (GC-MS) and UV-visible absorption, respectively, to analyse product formation. The electrocatalysis was first probed with all the reaction components in solution. In the absence of the natural electron source, NADPH (Supplementary Fig. S9a), the electrocatalytic reaction is driven by the electrode and requires either the diffusion of the enzyme or an electron mediator to/from the electrode surface (Supplementary Fig. S9b). Fig. 3a and b displays the electrocatalysis of P450$_{cam}$ and GR, respectively, studied by CV. The two CV signals per each enzyme correspond to the redox (non-catalytic) and catalytic conditions, namely, absence (P450$_{cam}$) and presence of the D-camphor substrate (P450$_{cam}$/camphor) in Fig. 3a (bottom panel), and absence (MV/GSSG) and presence of the enzyme (MV/GSSG/GR) in Fig. 3b (bottom panel). The voltametric signals obtained are concordant with previous literature: (1) P450$_{cam}$ presents a redox peak between -0.4 V and -0.6 V[34–36] with observed electrochemical current enhancement at -0.27 V in the presence of D-camphor. The redox mid-point shifts by ∼0.17 V as the enzyme goes from its substrate-free to substrate-bound form[37,38,39], as observed by the slightly more positive electrocatalytic signal, indicating catalytic transformation of camphor to 5-*exo*-hydroxycamphor (Fig. 3a, top panel)[36]. The shift in P450$_{cam}$ redox potential accompanies changes in the optical spectrum of the enzyme upon camphor binding due to the Fe(III) in the *heme* co-factor going from a high to low spin state (Supplementary Fig. S4)[40]. The increase in cathodic (negative) current coincides with the observed increase of enzyme apparent height in the STM image for the surface-absorbed P450$_{cam}$ (Fig. 1e). (2) The electrocatalytic reduction of GSSG to GSH by GR is mediated by methyl viologen dichloride (MV) as the electron mediator between the enzyme and the electrode surface (Fig. 3b, top panel)[41,42,43]. MV provides means to record the electrocatalytic activity of GR via the MV electrochemistry, since GR displays no redox signal within the aqueous electrochemical potential range. MV is the only species that displays redox chemistry at -0.68 V, whilst GSSG and GR displayed no redox chemistry in the working potential window (Supplementary Fig. S10b)[44]. The electrocatalysis of GR is a much faster reaction than the P450$_{cam}$ resulting in higher catalytic currents, which is supported by the much larger $k_{cat}$ turnover rate for GR[22,25].

The same electrocatalytic experiments were also conducted by pre-adsorbing the enzyme on the C8-Au electrode surface (Supplementary Fig. S9c) to confirm enzymatic activity under conditions similar to those used in the single-enzyme experiments. We observed very limited enzyme coverage of the electrode surface (Fig.1c-d) resulting in insufficient enzyme to produce detectable concentrations of reaction product in the short CV-based electrocatalytic assays. To overcome this limitation, chronoamperometric experiments were carried out to drive the electrocatalytic reaction over long time periods of 3 hours for GR and up to 7 hours for P450$_{cam}$. This was followed by the chemical quantification of product formation using GCMS[45]. The findings are summarised in Supplementary Figs. S11-S12 and provide support that the surface-adsorbed enzymes maintain enzyme catalytic activity.

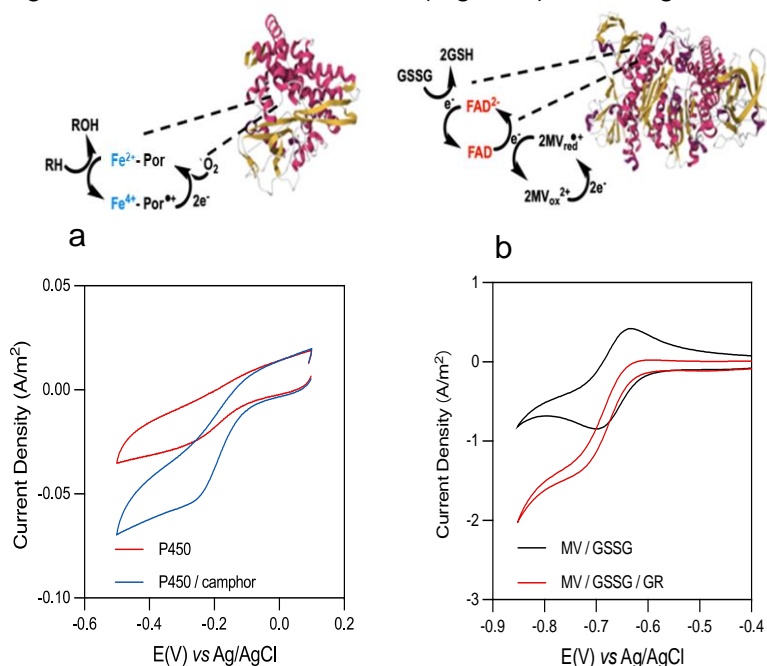

**Fig. 3**. **Solution phase electrocatalytic cyclic voltammetry (CV). a**, CVs of P450 in the absence (P450) and presence of camphor (P450 / camphor), using 0.1 M phosphate buffer (pH 7). **b,** CVs of GR in the presence of MV and GSSG only (MV / GSSG) and presence of MV, GSSG and GR (MV / GSSG / GR) using 0.1 M Tris-KCl (pH 7). The corresponding electrocatalytic reaction schemes are represented on the top panels. CVs were obtained using octanethiol-functionalised Au working electrode, Pt counter electrode and Ag/AgCl (3 M KCl) reference electrode. Scan rates for P450 and GR electrocatalysis were 5 mV s$^{-1}$ and 10 mV s$^{-1}$, respectively.

## Electrochemically Controlled Single-Enzyme Catalytic Junctions

Transient electrochemically controlled nanogaps were formed using an EC-STM configuration as described above over both the P450$_{cam}$-C8-Au and GR-C8-Au interfaces containing a low coverage of proteins scattered on the electrode surface (Fig. 1c-d, respectively). Sudden current jumps, as a result of individual enzyme trapping in the tunnelling nanojunction (Fig. 2a-b), were recorded at both oxidising and reducing applied potentials in the presence and absence of their corresponding enzymatic substrates and/or redox mediator. A reducing potential of -0.4 V was applied to the P450$_{cam}$-C8-Au interface in oxygenated aqueous buffer containing the substrate camphor, to induce the electrocatalytic enzymatic process. Molecular oxygen is a co-substrate in the catalysed oxidation of camphor by P450$_{cam}$. For GR-C8-Au, the substrate GSSG and electron mediator MV are added to the buffer at a reducing applied potential of -0.5 V. We define these as the *active conditions* whereby the enzymatic interfaces have shown to undergo electrocatalysis (Fig. 3). In contrast, we define various *inactive conditions* as: (1) when both interfaces are in the presence of all chemical components for electrocatalysis, but subjected to applied oxidising potentials of +0.1 V, far beyond where electron injection for electrocatalysis occurs, and (2), in the absence of their catalytic components, namely, substrate and/or redox mediator, at either reducing or oxidising potentials. These inactive conditions are control experiments where bulk electrocatalysis is supressed (Fig. 3). Two different types of protein trapping events (*blinks*) are observed during the lifetime of the tunnelling nanogaps, which we denote as *silent* (Fig. 4a, lower panel) and *switching* (Fig. 4b, lower panel). An algorithm in python was coded to automate the classification of all the collected blinks under the different conditions (1000–1500 traces per condition) into the two *silent* and *switching* categories (Supplementary Fig. S21). The *silent* blinks typically display a single conductance level with small fluctuations around the average protein conductance (Fig. 2), in line with values obtained in previous single-protein junction work[7]. In contrast, the *switching* blinks display a characteristic two-levels fluctuation of the conductance superimposed on top of the blink during the protein residence time in the nanojunction (Fig. 4b, lower panels for $P450_{cam}$-C8-Au (left) and GR-C8-Au (right)). A first visual inspection shows that the frequencies of the two-levels fluctuations across the two enzymes are strikingly different, with much higher values for GR. The latter shows a first correlation between the frequency of the observed 2-level fluctuation signal and the characteristic $k_{cat}$ of the enzyme. Similar fluctuating signals to those observed for GR were recently reported by Zhang *et al*. in a non-redox, catalytically fast ($k_{cat}$ = 7200 min$^{-1}$ [46]) Φ29-polymerase[5]. The silent blinks observed in the active conditions were compiled in 1D histograms and compared to previous analysis in the inactive conditions (Supplementary Fig. S19). Except for the GSSG substrate, similar conductance values around 2-3 x 10$^{-5}$ $G_0$ are observed for both $P450_{cam}$ and GR in the presence of substrate and/or mediator, indicating that substrate/mediator binding to the enzyme does not significantly impact the protein conductance measured during the single protein trapping. All switching signals recorded under active conditions were accumulated in 1D histograms with no selection for both $P450_{cam}$ and GR (Fig. 4c-d, respectively). A bimodal distribution is observed in both cases, where G1 (Fig. 4c-d, red curve) and G2 (Fig. 4c-d, blue curve) correspond to the averaged two levels of conductance arising from the conductance fluctuation in the switching blink. The conductance values for G1 and G2 are 1.7 x 10$^{-5}$ $G_0$ and 4.6 x 10$^{-5}$ $G_0$ for $P450_{cam}$ and 4.5 x 10$^{-5}$ $G_0$ and 1.5 x 10$^{-4}$ $G_0$ for GR, respectively, indicating a factor of 2.7 and 3.3 of conductance change between G1 and G2 for $P450_{cam}$ and GR, respectively. A comparison of G1 with previous conductance values (Fig. 2c-d) shows that the first conductance level matches the intrinsic protein conductance of P450 at off-resonant potentials from the *heme* redox centre, thus suggesting that camphor substrate binding does not significantly impact protein conductance. A slight 1.7 increase for GR conductance is only observed upon GSSG substrate binding.

The previous classification also enables analysis of the percentages of observed switching blinks out of the total number of blinks (silent + switching) collected at each experimental condition for both $P450_{cam}$ and GR (Fig. 5a-b, respectively). An abrupt increase in the number of observed switching signals is consistently observed under the active conditions for both enzymes, reaching ~30% of the total number of traces. A residual 5-7% of switching blinks is still recorded under inactive conditions, suggesting the occurrence of enzymatic events that do not lead to the enzymatic chemical conversion.

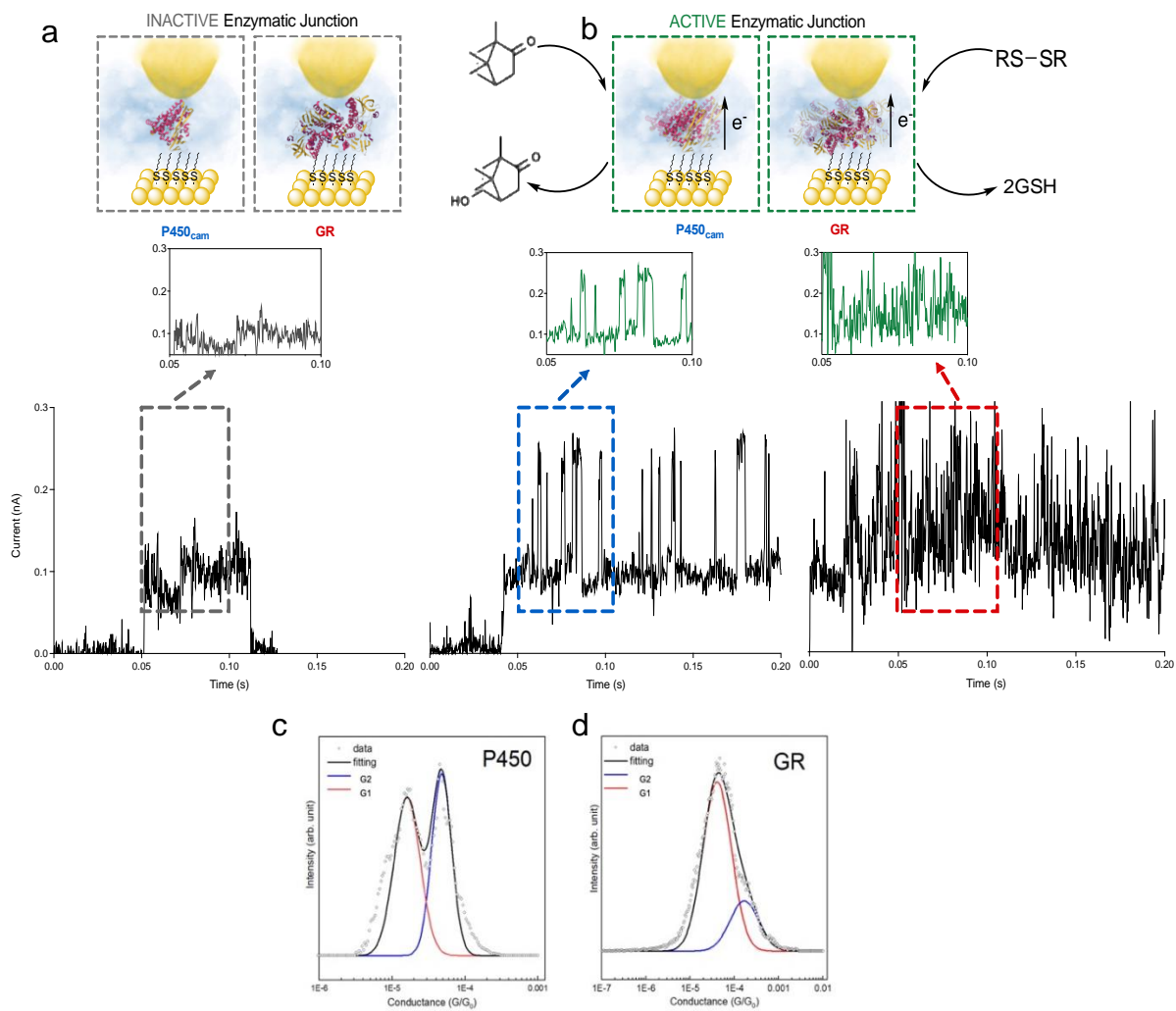

**Fig. 4**. **Single-protein conductance measurements under electrocatalytic conditions**. **a**, Current transients of an inactive single enzymatic junction obtained in the absence of substrates. **b**, Current transients of active P450-C8-Au (left) and GR-C8-Au (right) junctions displaying switching events in the presence of their substrates. **c**,**d**, 1D conductance histograms generated by accumulating switching traces of the enzymes in their active conditions: **c**, P450 in the presence of camphor at reducing potentials of -0.4 V and **b**, GR in the presence of GSSG and MV at reducing potentials of -0.5 V. Both experimental data (open circles) and its fitting (black) are represented, where red corresponds to the fitting of the first level of conductance (G1) and blue the second level of conductance (G2) resulting from the switching events. Conductance values for G1 and G2 are $1.7 \times 10^{-5}$ $G_0$ and $4.6 \times 10^{-5}$ $G_0$ for P450 and $4.5 \times 10^{-5}$ $G_0$ and $1.5 \times 10^{-5}$ $G_0$ for GR, respectively. Each histogram is built from a total of ~800-1000 current-time traces.

To seek a correlation between observed conductance switching and enzymatic activity, we first conducted an exhaustive frequency analysis of the conductance switching events using both a machine learning-based Gaussian Mixture Model (GMM)[47] and a Hidden Markov Model (HMM)[48] (see Supplementary section 3.2 for details). Both were able to successfully separate the two conductance levels (see Supplementary Figs. S22-23). Two different characteristic frequencies were calculated for each of the two enzymes under the active conditions, which we denote as the *bulk* and *single-enzyme* frequencies. The bulk frequency considers the number of times a switching event occurs in every switching trace averaged over the total protein time residence considering all signals (silent + switching) accumulated in long (>30 minutes) single-protein measurements (equation [1]). In contrast, the single-enzyme frequency considers only the occurrence of a switching event in the active (switching) traces, ruling out time averaging over the silent (non-active) traces (equation [2]). Note that the nature of the observed conductance fluctuations is stochastic and that the frequency values here refer to the conductance switching events expressed as turnover number per minute without implying periodicity. Typically, single-molecule enzymatic experiments measure the probability density or frequency of the stochastic occurrence of individual turnover events[49]. This has been shown in single-protein fluorescent experiments on a cholesterol oxidase enzyme, where each stochastic fluorescent blink is ascribed to an individual enzymatic turnover event, and the frequency was then calculated by considering the number of switching events per unit time[13,50]. Table 1 summarises the extracted frequencies from our single-protein junction experiments. The calculated bulk frequencies of the conductance fluctuation events align well with the reported kinetic $k_{cat}$ for P450$_{cam}$[22,23,24] and GR[25,26,27]. A few considerations on the obtained frequency numbers in Table 1 are: (1) previous single-molecule fluorescence microscopy experiments suggested that each switching event may not indicate a successful enzymatic turnover, but a conformational change that leads to an inactive state[4,13]. The conductance switching events in the single-protein experiments reflect the electrical transduction of all possible dynamics in the enzyme, whether they lead to product formation or not. This might likely result in overestimated bulk frequency values when compared to the intrinsic enzyme $k_{cat}$, which are solely based on successful product formation. The latter could be partially alleviated by considering the average (5-7%) switching dynamics observed in the single-protein experiments during the inactive conditions (Fig. 5) as an estimation of the average background single-protein dynamics leading to unsuccessful product formation in our experiments. Subtracting them from the bulk frequencies in the active conditions leads to a corrected value for the P450$_{cam}$ and GR bulk frequencies of 1790 and 12871 min$^{-1}$, respectively, which both lie well within the reported enzymatic turnover values for these enzymes (Table 1). (2) Kinetic constraints imposed by the nanogap are expected to occasionally lead to enzyme conformations with poor or no ability for the substate to get in and/or out of the enzyme's active site, resulting in a level of underestimation of the extracted single enzyme frequencies.

$$Bulk\ Freq = \frac{Number\ of\ Switching\ Events}{total\ protein\ trapping\ time} \qquad [1]$$

$$Single\ Enzyme\ Freq = \frac{Number\ of\ Switching\ Events}{total\ time\ of\ switching\ traces} \qquad [2]$$

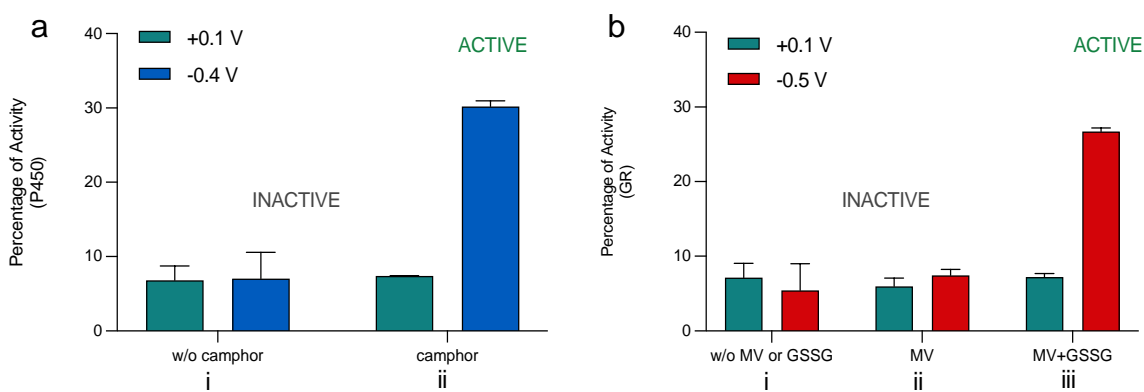

**Fig. 5**. **Percentage of activity in the current transient traces**. % of individual traces displaying conductance switching events under controlled electrochemical potentials for **a,** P450$_{cam}$-C8-Au at +0.1 V and –0.4 V (i) in the absence and (ii) in the presence of camphor, and for **b**, GR-C8-Au at +0.1 V and –0.5 V (i) in the absence of MV and GSSG, (ii) in the presence of MV only, and (iii) in the presence of MV and GSSG. The current transients were classified using a Python algorithm.

**Table 1.** Extracted bulk and single-enzyme turnover frequencies (min$^{-1}$) from the frequency of switching events on switching traces using Machine Learning Hidden Markov Model:

|  | **Bulk Freq (background corrected) (min$^{-1}$)** | **Single-Enzyme Freq (min$^{-1}$)** | **K$_{cat}$ (min$^{-1}$)** |
|---|---|---|---|
| **P450$_{cam}$** | 2573 (1790) | 6908 | 124-3960 [22,23,24] |
| **GR** | 16292 (12871) | 42943 | 12600-17500 [25,26,27] |

The analysis of bulk frequencies (Eq. [1]) provides a good foundation supporting the correlation between the observed single-protein conductance fluctuations and the real-time monitoring of the protein enzymatic activity. Equation [2] above describes the "local" or intrinsic single-enzyme frequency, which characterizes the enzymatic activity of a single enzyme within a population where dynamic activity disorder exists. We observe a ~3-fold increase for both enzymes when going from the bulk to the single-enzyme frequencies (Table 1), which reflects the significantly faster intrinsic enzyme activity not captured in the ensemble experiments. These single-enzyme catalytic rates have been previously reported for a horse radish peroxidase enzyme using fluorescence labels, showing a factor of 10 over the bulk characteristic turnover values[51]. Our non-labelled, electrically transduced single-protein enzymatic activity correlates with other single-molecule enzymatic turnover experiments in demonstrating that the overall frequency of an enzymatic reaction is a manifestation of dynamic disorder, where there is a fluctuation in rate constants due to a continuous, stochastic transition between on-off states of enzyme's activity[49]. This demonstrates the high heterogeneous nature of enzymatic catalysis and opens to a detailed study of the complex dynamics in redox enzymatic processes.

## Proposed mechanisms for the electrical transduction of enzymatic events

A mechanism has been long hypothesized where conductance switching in a single-enzyme junction is used as means to transduce and *in situ* monitor enzymatic catalysis at real-time[12]. There is substantial work reporting on conductance switching in single redox-active proteins using both high resolution EC-STM imaging[11] and single-protein junction measurements[7,52]. The telegraph-like fluctuations in the single-molecule current traces contain information about the dynamic changes of the protein redox state. This has been described in detail in previous single-molecule electrical characterization on a model redox Cu Azurin[52] as well as other synthetic redox molecules[60], where the fluctuations between conductance values for the oxidised and reduced states of the molecule were electrochemically driven and the reduced-to-oxidized population ratio fit a Lorentzian function that nicely predicts the protein redox mid-point potential at the population ratio 1:1[52]. The subtle change in protein conductance occurs when the energy of the protein redox state is brought into resonance with the Fermi energies of the two voltage-biased junction metal electrodes via the applied electrochemical "gating" potential. This process has been modelled at length by Kutnesov and Ulstrup[10,53], which describes the phenomenon as the opening of a new two-step sequential electron tunnelling channel where the charge transport is mediated by the redox centre (Fig. 6b). The manifestation of such charge transport mechanism in P450$_{cam}$ is shown in the potential dependence of the STM image apparent height (Fig. 1e), in line with the same analysis in other model redox proteins and redox cofactors[9,11]. When the potential applied coincides with the protein redox mid-point, the sequential tunnelling opens and protein conductance is enhanced resulting in a higher apparent height of the protein STM image[9,11,54]. How does the sequential tunnelling scenario relate to the catalytic cycle? In the above examples, the redox resonance conditions leading to sequential tunnelling are driven electrochemically by bringing the redox state between the Fermi energy of the two nanojunction electrodes[7,11,53,54]. With analogy to the latter, the redox enzyme undergoes subtle redox state changes of the active site co-factor during each catalytic cycle. In the single-protein electrocatalytic experiments, the conductance switching occurs at applied negative overpotentials with respect to the P450$_{cam}$ redox mid-point, which leaves the redox energy state off-resonance (Fig. 6a) leading to protein conductance mediated by electron tunnelling through the peptide structure of the protein[55-58]. During the P450$_{cam}$'s catalytic cycle, upon substrate and oxygen binding, the enzyme's cofactor *heme* is momentarily oxidised from reduced Fe(II)-Por to Fe(IV)-Por$^{•+}$ driven by the catalytic reaction (Fig. 3a (top panel) and Supplementary Fig. S25), bringing the redox state momentarily into resonance and opening the sequential electron tunnelling channel (Fig. 6b). This results in the observed conductance jumps that makes the switching trace (Fig. 6, bottom panel). The redox resonance condition leading to sequential tunnelling is now driven by the enzymatic process. After substrate enzymatic oxidation, the protein goes back to the reduced state recovering the off-resonance tunnelling conductance value (Fig. 6a). Fig. 6 schematically illustrates the mechanism for the momentary, chemically driven opening of the sequential tunnelling channel that accounts for the observed two-level conductance switching during the catalytic process. A more in-depth analysis of the potential-dependent STM images in Supplementary Fig. S20 shows the same switching behaviour when imaging the enzyme under active condition.

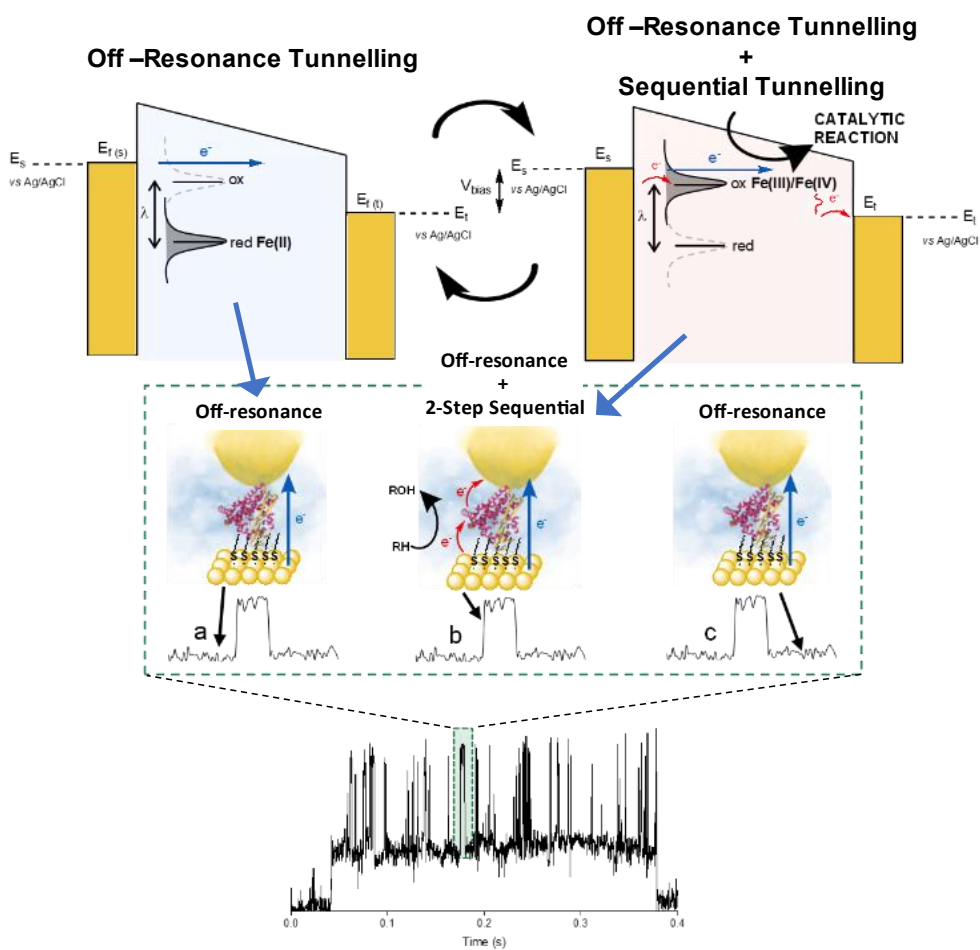

**Fig. 6**. **Proposed tunnelling mechanism for the observed two-level conductance**. At applied reducing potentials to the substrate, **a**, the *heme* centre is reduced to Fe(II) and the first level of conductance represents a non-catalytic off-resonant coherent tunnelling channel, where electrons tunnel through the peptide structure of the enzyme. **b**, During the catalytic reaction, the *heme* centre momentarily oxidises to Fe(III)/Fe(IV), bringing the redox state into resonance, which opens an additional 2-step sequential tunnelling channel and results in the observed second-level of conductance. After the reaction is complete, the *heme* centre is reduced again to Fe(II) and the conductance decreases back to the off-resonance tunnelling channel. E$_{f(s)}$ and E$_{f(t)}$ represents the fermi level of the substrate and tip, respectively, which are held at electrochemical potentials of -0.4 V and -0.3 V, respectively *vs* Ag/AgCl. V$_{bias}$ = +0.1 V.

This is a first attempt to monitor real-time single-protein enzymatic activity using a non-labelled, all-electrical detection method. Our results are in line with previous reports on single-enzyme activity achieved by exploiting optical detection approaches using modified enzymes with various fluorescence labels[13,14,16]. Our methodology opens to new routes to characterize single-molecule enzymatic processes via electrical transduction of individual unmodified enzymes trapped in nanoscale tunnelling nanogaps that can be currently mass-produced using present technology[59].

## Conclusions

Here we demonstrate real-time electrical transduction of redox enzymatic catalysis on individual non-labelled enzymes trapped in a nanoscale tunnelling gap. We exploit controllable, electrochemical STM-based nanoscale junctions to transiently trap individual P450$_{cam}$ and GR redox enzymes between two electrodes and characterise their charge transport characteristics while the electrocatalytic enzymatic reaction is undergoing. The transient tunnelling current flowing across the single-enzyme junction shows a sharp occurrence increase of characteristic 2-level conductance fluctuations (switching signal) when the enzyme is subjected to the active catalytic conditions, namely, under applied electrochemically reducing potentials and in the presence of their required enzymatic substrates and/or redox mediator. The conductance switching appear superimposing the typical single tunnelling conductance trace (silent signal) typically observed during the formation of a single-protein junction under non-catalytic conditions. We experimentally demonstrate that the nature of the observed conductance switching events is driven by the dynamic change in the enzyme's redox state as it undergoes catalytic reaction; during a catalytic cycle, the *heme* active co-factor is momentarily oxidised which aligns the enzyme's redox energy level with the Fermi levels of the junction electrodes, opening an additional redox-mediated sequential tunnelling channel that transiently enhances the protein junction conductance. We used two different machine learning-based methods to perform an exhaustive classification of the silent and switching signals and extracted the frequencies for the observed switching events. Two different catalytic rates have been extracted from the single-enzyme assays for both enzymes: a *bulk frequency*, which correlates well with reported $k_{cat}$ determined from standard biochemical assays, and a *single-enzyme frequency*, which quantifies the single-enzyme characteristic turnover and highlights the heterogeneity nature of redox enzymatic process. These results demonstrate the feasibility of monitoring enzymatic reactions in real-time by electrically transducing catalytic events in a single-protein junction made of an unmodified enzyme, thus providing a new biophysical platform to study single-enzyme catalysis and opening to new avenues for label-free detection in biosensing.

## Methods

### Protein expression and purification

Recombinant P450cam was produced in *E.coli* TunerTM (DE3) competent cells (Novogen) with an N-terminal His-Tag (Supplementary Fig. S1). Cells were harvested by centrifugation (4200 rpm, 20 min, 4°C). The pellet was resuspended in a minimum volume of Buffer A (20 mM Tris buffer, 100 mM NaCl, 20 mM imidazole, 10% glycerol) with the addition of dithiothreitol (DTT) (final conc. 0.5 mM), Pepstatin A (final conc. 1 µg mL-1 per), DNAse A (final conc. 0.2 mg mL-1) and Complete™ protease inhibitor cocktail (1 tablet per 2 L of culture, from Sigma). Cell lysis was carried out on a cell disruptor IXT4A (Constant System LTD) with a pressure of 25 kPa, 4° C. The lysate was centrifuged (18,000 rpm) for 42 min at 4°C. The lysate was applied to a 5 mL His-TrapTM Fast Flow nickel affinity column (GE) equilibrated with Buffer A and connected to ÄKTA Pure Chromatography System (GE). His tagged proteins were eluted with Buffer B (20 mM Tris buffer, 100 mM NaCl, 200 mM imidazole, 10% glycerol). Fractions containing the protein of interest were concentrated to a volume less than 5 mL and injected onto a HiLoadTM 16/600 Superdex 200pg size exclusion chromatography column (GE) equilibrated with Buffer C (20 mM Tris buffer, 100 mM NaCl, 10% glycerol). The purification was monitored at 280 nm and 420 nm (Fig. S2). The proteins were eluted with Buffer C and fractions with *heme* incorporated protein were analysed by SDS-PAGE gel (Fig. S2) and were then concentrated and mixed with an equal volume of Buffer D (20 mM Tris buffer, 100 mM NaCl, 30% glycerol).
GR was bought from Sigma Aldrich.

## Preparation of Au(111)/Enzyme interfaces
Au(111) (MaTecK) was placed in a piranha solution (30% w/w hydrogen peroxide to concentrated sulphuric acid in a 1:3 ratio) for 4 minutes then rinsed with Milli-Q water. Next, the electrode was electrochemically oxidised in 0.1 M sulphuric acid for ~1.5 minutes and immersed in 0.1 M hydrochloric acid to dissolve the oxide layer for 20 minutes. Finally, the electrode was annealed in a hydrogen flame for 6 minutes and cooled under nitrogen gas for 2 minutes. Clean Au(111) substrates were placed in 1-octanethiol (1 mM) solutions using ethanol as a solvent for 18 hours. Then the electrodes were thoroughly rinsed with ethanol, then Milli-Q water and dried under nitrogen gas. P450$_{cam}$ / GR (20 μM) were incubated on the 1-octanethiol functionalised Au(111) for 2 hours at 4°C, then rinsed with Milli-Q water.

## Cyclic Voltammetry and Electrocatalysis
Measurements were carried out using a Metrohm-Autolab (the Netherlands) run on the NOVA software using a three-electrode configuration: 1.6 mm diameter polycrystalline Au working electrode, a long platinum wire counter electrode and a Ag/AgCl (3 M NaCl) reference electrode.
### *P450$_{cam}$*
The electrochemical/electrocatalytic measurements were conducted in an electrolyte containing P450$_{cam}$ (20 μM) and D-camphor (1 mM) in phosphate buffer (0.1 M, pH 7) as the supporting electrolyte. The potential voltage applied to the cell was swept from +0.1 V to -0.50 V at 0.005 V s$^{-1}$. Control experiments were carried out in the absence of both P450$_{cam}$ and camphor, and in the absence of camphor only.
### *GR*
The electrochemical/electrocatalytic measurements were conducted in an electrolyte containing glutathione reductase (4.4 μM), methyl viologen dichloride (1 mM) and GSSG (10 mM) in Tris-KCl (0.1 M, pH 7) as the supporting electrolyte. The potential voltage applied to the cell was swept from 0V to -0.85V at 0.01 V s$^{-1}$. Control experiments were carried out in the absence of GR and in the absence of both GR and GSSG.

## EC-STM Imaging and Nanogap formation
All electrochemical STM experiments were conducted in a Scanning Tunnelling Microscope (Bruker, USA) controlled by a Nanoscope V electronics run by Nanoscope software v9.7. Current-time measurements were conducted using a LabVIEW interface controlled by homemade LabVIEW codes. Electrochemical STM cells were cleaned using piranha before use. Surface-absorbed enzymes on the thiol-functionalised Au(111) substrates were mounted into a clean teflon cell. Phosphate buffer (0.1 M, pH 7) for P450$_{cam}$ or Tris-KCl (0.1 M, pH 7) for GR was used as the supporting electrolytes. All Au tips were electrochemically etched in 50:50 ethanol:HCl, functionalised with 1-octanethiol (1 mM, 1 hour), coated with apiezon wax and placed into the tip holder of the STM scanner equipped with a homemade 10 nA V$^{-1}$ current amplifier. The scanner and cell were mounted onto the STM where Au(111) at Au STM probe were the working electrodes, a Pt acts as the counter electrode and a miniaturised Ag/AgCl (3 M KCl) as the reference electrode.
Electrochemical STM images were obtained by scanning the tip in the XY plane of the substrates using a sample-to-tip voltage bias voltage of +0.1 V, current setpoint 1 nA and electrochemical potential of +0.1 V. P450-C8-Au was imaged in the presence and absence of camphor at different electrochemical potentials between -0.1 V and -0.45 V. The current-height of the individual P450$_{cam}$ enzymes were analysed from electrochemical STM images using Gwyddion software.

Electrochemical STM-based nanogap measurements were carried out using a sample-to-tip bias voltage of +0.1 V and current setpoint of ~40 pA. For P450$_{cam}$ junctions, D-camphor (250 μM) was added to the electrochemical cell and the sample potential was held at +0.1V and -0.4 V. For GR junctions, methyl viologen dichloride (157 μM) and GSSG (1527 μM) were added into the cell respectively, and the sample potential was held at +0.1 V and -0.5 V. In all cases, the STM probe potential was always +100 mV with respect to the sample potential. Thousands of blinks were collected per each experimental condition. Categorisation of the blinks are described in the Supplementary Information section 3.1.

## Acknowledgements
This work has been mostly supported by the European Research Council (ERC) under the European Union Horizon 2020 research and innovation program (Grant Agreement ERC Fields4CAT-772391) and from the UKRI-BBSRC BB/X002810/1 project. ID-P thanks King's College London for providing starting funds which partially supported this project. ACA thanks to the support of the Spanish MICINN with the fellowship RYC2021-031001-I and from the Max Planck Society. NK and JP thank NMES King's College London for financial support. We thank Oscar Ayrton assistance collecting GCMS data.

## Ethics declaration
All authors declare no competing interest.

## Supplementary Information
Section 1: General methods
Section 2: Single-enzyme conductance analysis
Section 3: Categorization and analysis of blinking traces
Section 4: Enzyme catalytic traces
Supplementary Figures S1-26.